\documentclass[a4paper]{jpconf}
\usepackage{graphicx}
\usepackage{xspace}

\def\gcm2{g/cm$^2$}

\def\knee{{\sl knee}\xspace}

\def\fref#1{Fig.\,\ref{#1}\xspace}
\def\and{\&\xspace}
\def\xx{\vspace*{-0mm}}

\def\1{$^1$}
\def\2{$^2$}
\def\3{$^3$}
\def\4{$^4$}
\def\5{$^5$}
\def\6{$^6$}
\def\7{$^7$}
\def\8{$^8$}
\def\9{$^9$}
\def\tor{$^{10}$}
\def\wup{$^{11}$}
\def\bn{$^{12}$}
\def\ldz{$^{13}$}
\def\k{$^,$}

\begin{document}
\title{Results from the KASCADE, KASCADE-Grande, and LOPES experiments}

\author{\begin{minipage}{0.82\textwidth}J R H\"orandel\1,
W D Apel\2, F Badea\2, L B\"ahren\3, K Bekk\2, A Bercuci\4, M Bertaina\5, P L
Biermann\6, J  Bl\"umer\1\k\2 , H  Bozdog\2, I M  Brancus\4, M
Br\"uggemann\7, P   Buchholz\7, S  Buitink\8, H  Butcher\3, A  Chiavassa\5, K
Daumiller\2, A G  de Bruyn\3, C M  de Vos\3, F  Di Pierro\5, P  Doll\2, R
Engel\2, J  Engler\2, H   Falcke\3\k\8, H  Gemmeke\9, P L  Ghia\tor, H -J
Gils\2, R  Glasstetter\wup, C   Grupen\7, D  Heck\2, A  Haungs\2, A
Horneffer\8, T  Huege\2, K -H  Kampert\wup, G W  Kant\3, H O  Klages\2, U
Klein\bn, Y  Kolotaev\7, Y   Koopman\3, O  Kr\"omer\9, J  Kuijpers\8, S
Lafebre\8, G  Maier\2, H J  Mathes\2, H J  Mayer\2, J  Milke\2, B  Mitrica\4, C
Morello\tor, M  M\"uller\2, G  Navarra\5, S   Nehls\2, A   Nigl\8, R
Obenland\2, J  Oehlschl\"ager\2, S  Ostapchenko\2, S  Over\7, H J  Pepping\3, M
Petcu\4, J  Petrovic\8, T  Pierog\2, S  Plewnia\2, H  Rebel\2, A   Risse\ldz, M
Roth\2, H  Schieler\2, G  Schoonderbeek\3, O  Sima\4, M St\"umpert\1, G
Toma\4, G C  Trinchero\tor, H  Ulrich\2, J  van Buren\2, W van Capellen\3, W
Walkowiak\7, A Weindl\2, S  Wijnholds\3, J  Wochele\2, J  Zabierowski\ldz, J A
Zensus\bn, and D  Zimmermann\7\end{minipage}}

\ifnum 1=2
\address{\1 University of Karlsruhe, Institute for Experimental Nuclear
         Physics, P.O. 3640, 76021 Karlsruhe, Germany}
\address{\2 Institut f\"ur Kernphysik, Forschungszentrum Karlsruhe, Karlsruhe,
         Germany}
\address{\3 ASTRON Dwingeloo, The Netherlands}
\address{\4 NIPNE Bucharest, Romania}
\address{\5 Departimento di Fisica Generale dell Universit`a Torino, Torino,
        Italy}
\address{\6 Max-Planck-Institut f\"ur Radioastronomie, Bonn, Germany}
\address{\7 Fachbereich Physik, University of Siegen, Germany}
\address{\8 Department of Astrophysics, Radboud University Nijmegen, The
         Netherlands}
\address{\9 IPE, Forschungszentrum Karlsruhe, Karlsruhe, Germany}
\address{$^{10}$Instituto di Fisica dello Spazio Interplanetario INAF, Torino,
        Italy}
\address{$^{11}$Fachbereich Physik, University of Wuppertal, Germany}
\address{$^{12}$Radioastronomisches Institut, University of Bonn, Germany}
\address{$^{13}$Soltan Institute for Nuclear Studies, Lodz, Poland}
\else
\address{\1 University of Karlsruhe, Institute for Experimental Nuclear
         Physics, P.O. 3640, 76021 Karlsruhe, Germany;
         \2 Institut f\"ur Kernphysik, Forschungszentrum Karlsruhe, Germany;\\
         \3 ASTRON Dwingeloo, The Netherlands;
         \4 NIPNE Bucharest, Romania;
	 \5 Dipartimento di Fisica Generale dell' Universit\`a  di Torino,
	    Italy;
         \6 Max-Planck-Institut f\"ur Radioastronomie, Bonn, Germany;
         \7 Fachbereich Physik, University of Siegen, Germany;
         \8 Department of Astrophysics, Radboud University Nijmegen, The
         Netherlands;
         \9 IPE, Forschungszentrum Karlsruhe, Germany;
	 $^{10}$ Istituto di Fisica dello Spazio Interplanetario INAF, Torino,
	         Italy;
         $^{11}$Fachbereich Physik, University of Wuppertal, Germany;
         $^{12}$Radioastronomisches Institut, University of Bonn, Germany;
         $^{13}$Soltan Institute for Nuclear Studies, Lodz, Poland}
\fi

\ead{hoerandel@ik.fzk.de}

\begin{abstract}
The origin of high-energy cosmic rays in the energy range from $10^{14}$ to $10^{18}$~eV is explored with the KASCADE and KASCADE-Grande experiments.
Radio signals from air showers are measured with the LOPES experiment.
An overview on results is given.
\end{abstract} \vspace*{-9mm}

\section{Introduction}

One of the most remarkable structures in the energy spectrum of cosmic rays is
a change of the spectral index $\gamma$ of the power law $dN/dE\propto
E^\gamma$ at an energy of about 4~PeV, the so called \knee
\cite{naganowatson,pg,haungsrebelroth}.  The origin of the \knee has not been
resolved yet and a convincing explanation of the \knee structure is thought to
be a corner stone in understanding the origin of galactic cosmic rays.  In the
literature various reasons for the \knee are discussed, being related to the
acceleration and propagation processes of cosmic rays as well as to
interactions in interstellar space or the Earth's atmosphere
\cite{origin,ecrsreview}.

The steeply falling energy spectrum requires large detection areas at high
energies.  The largest balloon-borne experiment with single-element resolution
(TRACER, 5~m$^2$\,sr) reaches energies of a few $10^{14}$~eV \cite{tracer05}.
To study higher energies, experiments covering several $10^4$~m$^2$ and exposure
times exceeding several years are necessary, which, at present, can only be
realized in ground-based installations. They measure the secondary products
generated by high-energy cosmic-ray particles in the atmosphere -- the
extensive air showers.  The challenge of these investigations is to reveal the
properties of the shower inducing primary particle behind an absorber -- the
atmosphere -- with a total thickness at sea level corresponding to 11 hadronic
interaction lengths or 30 radiation lengths.

One of the most advanced experiments in the energy range from 
$10^{13}$~eV to $10^{17}$~eV is the experiment KASCADE ("KArlsruhe
Shower Core and Array DEtector") \cite{kascadenim}.  It is continuously
operating since 1996, detecting simultaneously the three main components of air
showers.  A $200\times 200$~m$^2$ scintillator array measures the
electromagnetic and muonic components ($E_\mu>0.23$~GeV).  The central detector
system combines a large hadron calorimeter, measuring the energy, as well as
point and angle of incidence for hadrons with energies $E_h>50$~GeV
\cite{kalonim}, with several muon detection systems ($E_\mu>0.49$, 2.4~GeV)
\cite{mwpcnim}.  In addition, high-energy muons are measured by an underground
muon tracking detector equipped with limited streamer tubes ($E_\mu>0.8$~GeV)
\cite{mtdnim}.

Results of the KASCADE experiment are summarized in
Sects.\,\ref{wwsect} - \ref{especsect}. 
In several astrophysical models a transition from galactic to extragalactic
cosmic radiation is expected at energies from $10^{17}$ to $10^{18}$~eV.  A
particularity in the energy spectrum in this range is the second \knee at about
400~PeV \cite{naganowatson,pg}.
The KASCADE-Grande experiment operates in this energy region, recent results
will be reported in Sect.\,\ref{grandesect}.
An alternative method to investigate high-energy cosmic rays is the detection
of radio signals from air showers, the status of the LOPES experiment will be
discussed in Sect.\,\ref{radiosect}.

\section{High-energy interactions and air showers}\label{wwsect}


Addressing astrophysical questions with air-shower data necessitates the
understanding of high-energy interactions in the atmosphere. 
The interpretation of properties of primary radiation derived from air-shower
measurements depends on the understanding of the complex processes during the
cascade development.  Recent investigations indicate inconsistencies in the
interpretation of air shower data \cite{rothnn,chicagoknee,pg}.  Thus, one of
the goals of KASCADE is to investigate high-energy interactions and to improve
contemporary models to describe such processes.
The program CORSIKA \cite{corsika} is applied to calculate the development of
extensive air showers. It contains several models to describe hadronic
interactions at low and high energies.  Their predictions are compared with
experimental results in order to check their correctness.

Studies of the shower development in the atmosphere have been performed with
the multi-detector set-up and interaction models have been improved
\cite{wwtestjpg,kascadelateral,rissejpg,kascadeabslength,hadrisvhecricern,annaprd,mayerapp}.
A valuable tool to test high-energy interaction models are correlations between
different shower components \cite{jenskrakow,jenspune}.  
A couple of years ago some models like SIBYLL\,1.6, DPMJET\,2.5, or NEXUS\,2
failed to describe the measurements of particular correlations. On the other
hand, for contemporary models like QGSJET\,01, SIBYLL\,2.1, or DPMJET~2.55, the
KASCADE measurements are compatible with predictions for various correlations
between the electromagnetic, muonic, and hadronic components, i.e.\ the
measurements are bracketed by the extreme assumptions of primary protons and
iron nuclei \cite{jenskrakow,jenspune}.
While in previous analyses  pure proton or iron compositions have been assumed
as extreme cases, at present, more detailed analyses are performed
\cite{jenspune,jrhtorun}. They take into account the spectra for elemental
groups as obtained from investigations of the electromagnetic and muonic
components (see below) and reveal deviations between measurements and
simulations for the hadronic component of the order of 10\% to 20\%.


In conclusion, the models QGSJET\,01 \cite{qgsjet}, SIBYLL\,2.1 \cite{sibyll21},
and DPMJET\,2.55 \cite{dpmjet} seem to be the most reliable models to describe
high-energy hadronic interactions.  However, also they are not able to describe
the data fully consistently \cite{ulrichapp}.  This illustrates the possibility
of experiments like KASCADE, which are able to study details of high-energy
interactions in the atmosphere and indicates the progress made in this field
during the last decade.  At the same time, this stimulates new efforts, with
the objective to improve the present interaction models. Examples are new
theoretical concepts included in the QGSJET model \cite{qgsjet2}, or the
investigations of dedicated changes of physical parameters like the inelastic
proton-proton cross-section or the inelasticity of hadronic interactions on
air-shower observables \cite{wq,isvhecri04kascadewq}.  The latter indicate that
variations of interaction parameters within the error bounds of accelerator
measurements yield significant and measurable changes in the air shower
development \cite{kascadewqpune}.

\section{Cosmic-ray anisotropy}
\begin{figure}[t]
 \includegraphics[width=0.49\textwidth]{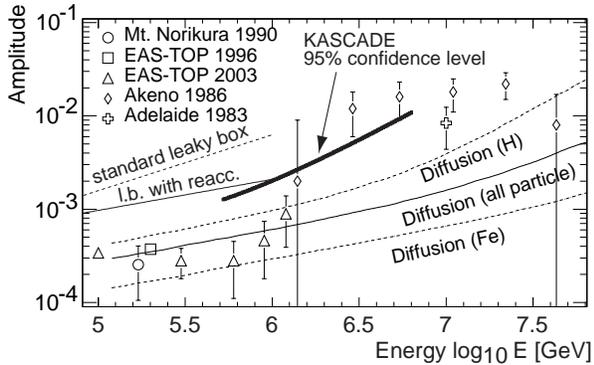}\hspace*{\fill}
 \begin{minipage}[b]{0.47\textwidth}	 
  \caption{Rayleigh amplitudes as function of energy for various
	   experiments, for references see \cite{kascade-aniso}.  Additionally,
	   model predictions for Leaky Box models \cite{ptuskinaniso} and a
	   diffusion model \cite{candiaaniso} are shown. For the latter, the
	   lines indicate the expected anisotropy for primary protons, iron
	   nuclei, and all particles.}
  \label{aniso}	  
 \end{minipage}	  
\end{figure}

Supernova remnants, such as Cassiopeia A, have been observed in electromagnetic
radiation in a wide energy range up to TeV-energies.  Calculations indicate
that the observed multi-wavelength spectra are consistent with the acceleration
of cosmic-ray electrons and hadrons in supernova remnants \cite{berezhko-casa}.
Recent observations by the H.E.S.S. experiment reveal a shell structure of the
supernova remnant RXJ-1713 and an energy spectrum of $\gamma$-rays $\propto
E^{-2.2}$ in agreement with the idea of particle acceleration in a shock front
\cite{hesssnr}.

Also, of great interest is to study the arrival direction of charged cosmic
rays to search for potential point sources.  The arrival directions of showers
with energies above 0.3~PeV covering a region from $10^\circ$ to $80^\circ$
declination have been investigated with KASCADE \cite{kascade-points}.  No
significant excess has been observed neither for all showers, nor for muon-poor
events.  The analysis has been deepened by investigating a narrow band
($\pm1.5^\circ$) around the Galactic plane. Also circular regions around 52
supernova remnants and 10 TeV-$\gamma$-ray sources have been studied.  None of
the searches provided a hint for a point source, neither by taking into account
all events, nor selecting muon-poor showers only.  Upper limits for the fluxes
from point like sources are determined to be around
$10^{-10}$~m$^{-2}$s$^{-1}$. In addition, no clustering of the arrival
direction for showers with primary energies above 80~PeV is visible.

While the search for point sources is related to the investigation of
cosmic-ray acceleration sites, the large scale anisotropy is expected to reveal
properties of the cosmic-ray propagation.  The Rayleigh formalism is applied to
the right ascension distribution of extensive air showers measured by KASCADE
\cite{kascade-aniso}.  No hints of anisotropy are visible in the energy range
from 0.7 to 6~PeV. This accounts for all showers, as well as for subsets
containing showers induced by predominantly light or heavy primary particles.
Upper limits for Rayleigh amplitudes are shown in \fref{aniso}.
The increase of the amplitudes as function of energy is predicted by
calculations using a diffusion model to describe the cosmic-ray propagation in
the Galaxy \cite{candiaaniso}.  This indicates that leakage from the Galaxy
plays an important part during cosmic-ray propagation and most likely, the
leakage is also (partly) responsible for the origin of the \knee. On the other
hand, simple Leaky-Box models seem to be ruled out by the measurements
\cite{kascade-aniso,maierflorenz,aspenphen}.


\section{Energy spectra and mass composition of cosmic rays}\label{especsect}
\begin{figure}[t]
 \includegraphics[width=0.49\textwidth]{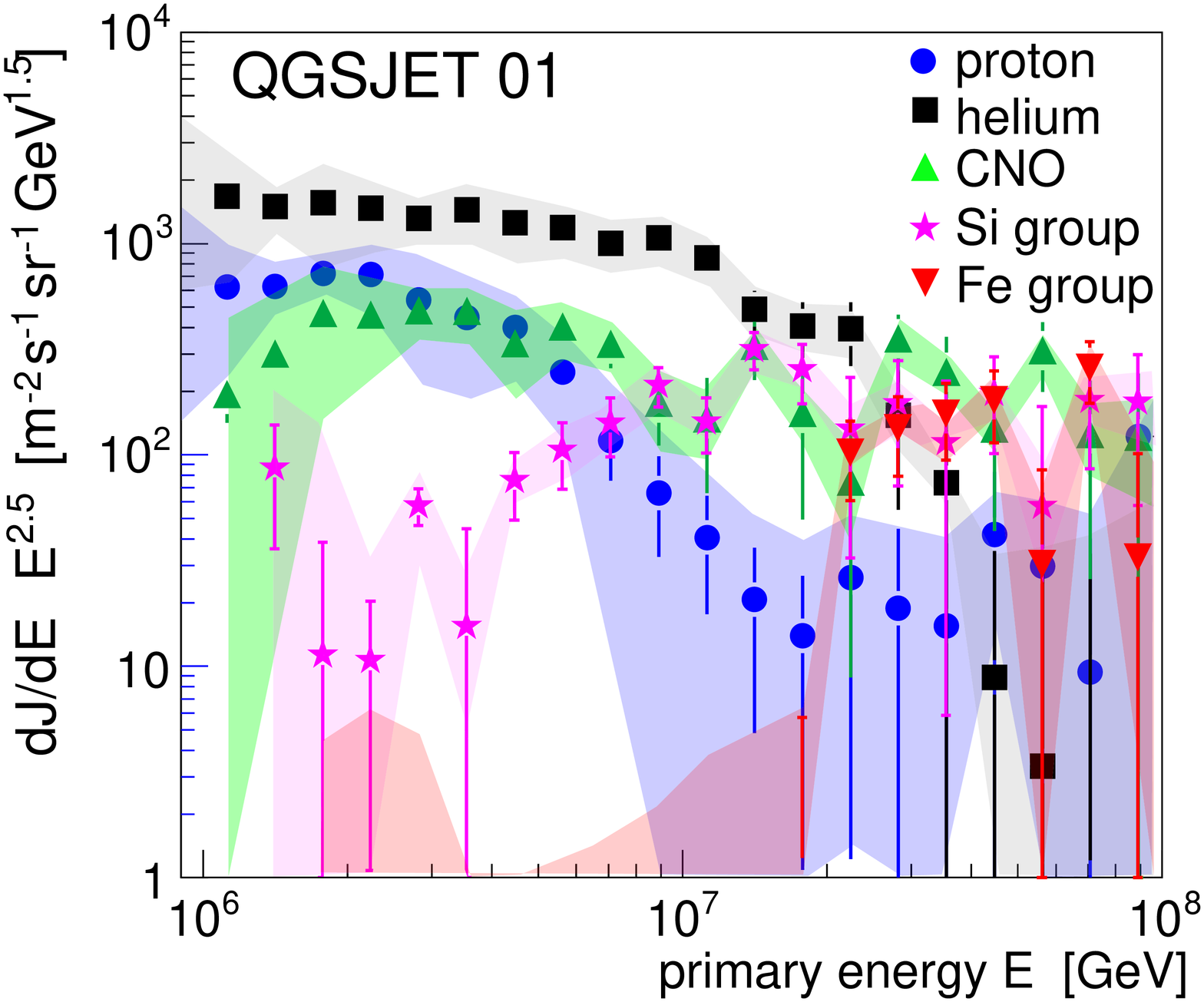}\hspace*{\fill}
 \includegraphics[width=0.49\textwidth]{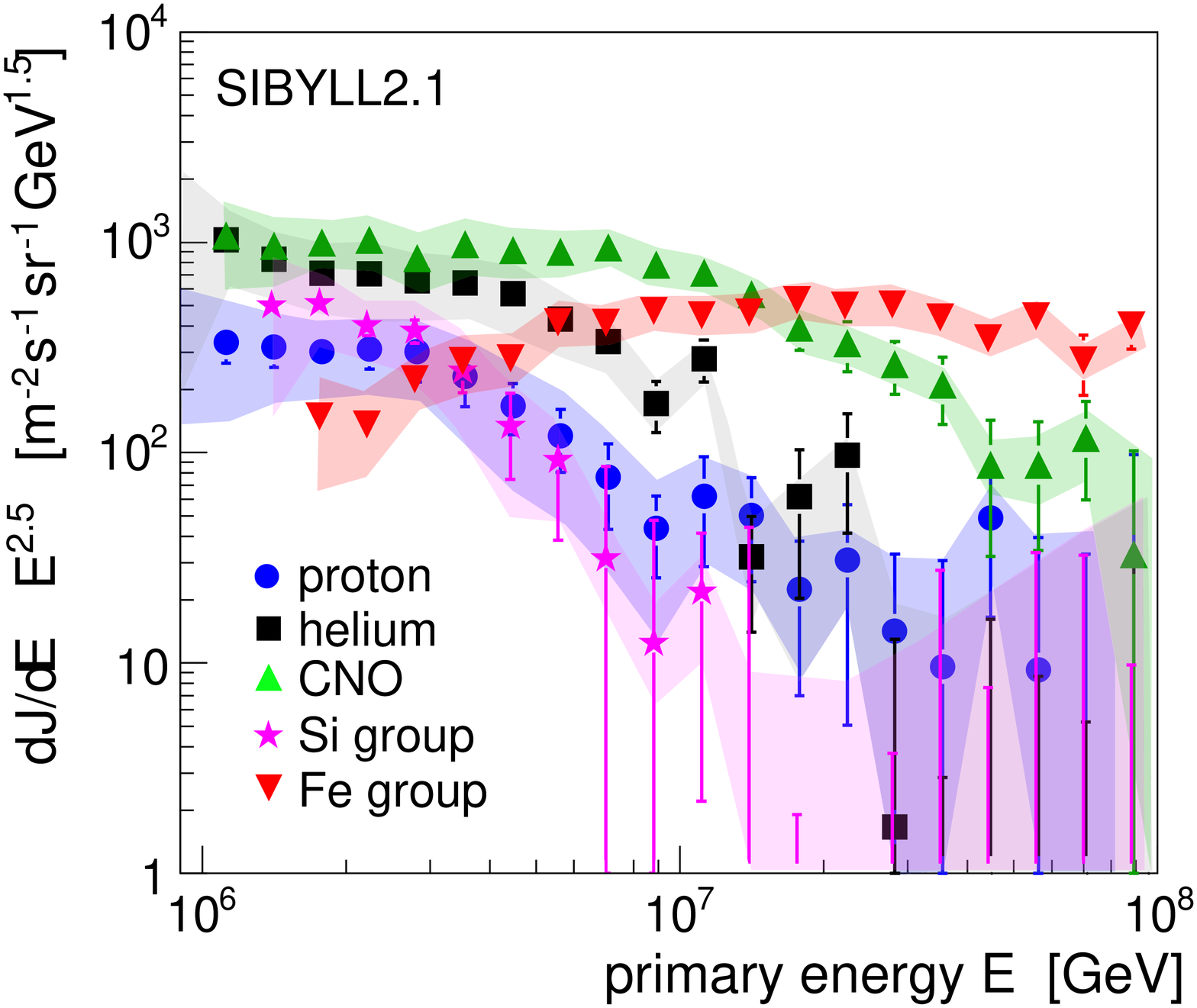} \xx
 \caption{Energy spectra for five cosmic-ray mass groups for measurements
	  interpreted with two different models to describe high-energy
	  hadronic processes in the atmosphere: QGSJET (left) and SIBYLL
	  (right) \cite{ulrichapp}. The bands represent systematic errors.}
 \label{holger}	  
\end{figure}

The main objective of KASCADE is to determine the energy spectra and mass
composition of cosmic rays. The problem has been approached from various points
of view.  It could be shown that a \knee exists in all three main shower
components, i.e.\ electrons, muons, and hadrons at energies $\approx 4-5$~PeV
\cite{allknee}.  The primary energy spectrum could be established based on the
electromagnetic and muonic \cite{glasstetterslc} as well as the hadronic and
muonic components \cite{hknie}.  An analysis of muon densities showed that the
\knee in the all-particle spectrum is caused by a suppression of light elements
\cite{muden}.  Analyses of the electromagnetic and muonic shower components
\cite{weber}, the hadronic and muonic components \cite{kascadehm}, as well as
various combinations of them \cite{rothnn} indicate an increase of the mean
logarithmic mass of cosmic rays as function of energy in the \knee region.  The
longitudinal development of the muonic shower component is studied with the
muon tracking detector of the KASCADE-Grande experiment \cite{buettner}.  The
measured flux of unaccompanied hadrons at ground level has been used to derive
the spectrum of primary protons \cite{kascadesh}. The resulting flux follows a
single power law in the energy range from 100~GeV to 1~PeV and is compatible
with direct measurements. 

An advanced analysis is founded on the measurement of the electromagnetic and
muonic shower components \cite{ulrichapp}.  It is based on the
deconvolution of a two-dimensional electron muon number distribution.
Unfolding is performed using two hadronic interaction models (QGSJET\,01 and
SIBYLL\,2.1) to interpret the data. The spectra obtained for five elemental
groups are displayed in \fref{holger}. They exhibit sequential cut-offs in the
flux for the light elements.  For both models a depression is visible for
protons around 3 to 4~PeV and at higher energies for helium nuclei.  The
systematic differences in flux for the spectra derived with QGSJET and SIBYLL
amount to a factor of about two to three.  The silicon and iron groups show a
rather unexpected behavior for both models.  The increase of the flux for both
groups (QGSJET) and the early cut-off for the silicon group (SIBYLL) is not
compatible with contemporary astrophysical models.  The discrepancies are
attributed to the fact that none of the models is able to describe the observed
data set in the whole energy range consistently \cite{ulrichapp}. 

Despite of the discrepancies, the spectra compare well to the results obtained
by the EAS-Top experiment \cite{eastopspec} and extend the results of direct measurements to
high energies \cite{aspenreview}.  Considering the energy range above 10~GeV,
at least a qualitative picture of the energy spectra for individual mass groups
emerges: the spectra seem to be compatible with power laws with a cut-off at
high energies.  The cut-off behavior indicated by the measurements is reflected
by theoretical considerations taking into account the maximum energy attained
during acceleration in supernova remnants \cite{sveshnikova} or diffusive
propagation of cosmic rays in the Galaxy \cite{kalmykov}. 

\section{Towards the second knee and the transition to extragalactic cosmic
         rays}\label{grandesect}

Energy spectra have been reconstructed with KASCADE data up to energies of
100~PeV. At these energies statistical errors start to dominate the overall
error. To improve this situation, the experiment has been enlarged.  Covering
an area of 0.5~km$^2$, 37 detector stations, containing 10~m$^2$ of plastic
scintillators each, have been installed to extend the original KASCADE set-up
\cite{grande}.  Regular measurements with this new array and the original
KASCADE detectors, forming the KASCADE-Grande experiment, are performed since
summer 2003 \cite{chiavassapune}.  In parallel,  a flash ADC system is being
developed to measure the time structure of air showers \cite{brueggemannpune}.
The objective is to reconstruct energy spectra for groups of elements up to
$10^{18}$~eV \cite{haungsaspen}, covering the energy region of the second
\knee, where the galactic cosmic ray spectrum is expected to end
\cite{aspenphen}.

\begin{figure}[t]
 \includegraphics[width=0.49\textwidth]{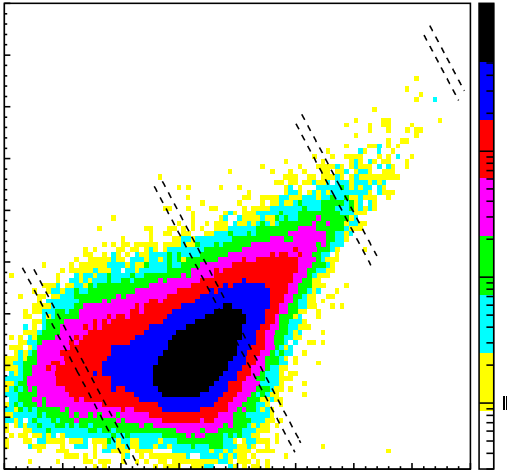} \hspace*{\fill}
 \begin{minipage}[b]{0.47\textwidth}	 
 \caption{
	  Reconstructed number of electrons as function of the number of muons.
	  The dashed lines indicate estimates for the primary energy for
	  showers with zenith angles of 0$^\circ$ and 18$^\circ$.
	  Parallel to the lines light elements are at the top and heavy 
	  elements at the bottom of the distribution
	  \cite{glasstetterpune}.}
 \label{grande}	  
 \end{minipage}
\end{figure}

First analyses extend the lateral distributions of electrons and muons up to
600~m \cite{glasstetterpune,vanburenpune}.  
A measured two-dimensional shower size spectrum is shown in \fref{grande},
the number of electrons is plotted as function of the number of muons.
To guide the reader, the dashed lines indicate an estimated primary energy.
One recognizes that already now with this data set, based on one year of
measurements, energies close to $10^{18}$~eV are reached. It is planned to
conduct an unfolding analysis, similar to the one described above, and reveal
the energy spectra for groups of elements up to $10^{18}$~eV.

\section{Radio emission from air showers}\label{radiosect}
\begin{figure}[t]
 \includegraphics[width=0.49\textwidth]{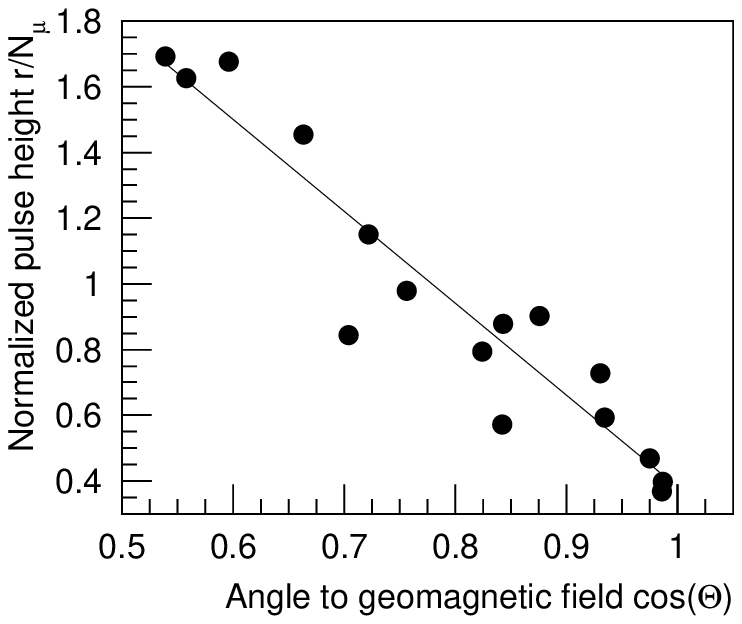}\hspace*{\fill}
 \includegraphics[width=0.49\textwidth]{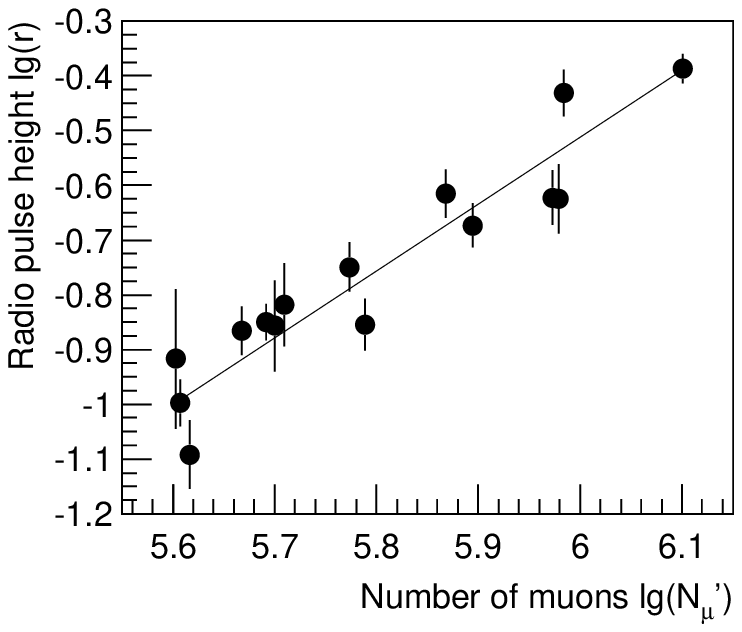}\xx
 \caption{Dependence of the measured radio signal yield on the angle between
	  the geomagnetic field and the shower axis ({\sl left}) and the number
	  of muons in the shower ({\sl right}).}
 \label{lopes}	  
\end{figure}

Radio emission from air showers is known since the 1960ies \cite{jelleynature}.
Most likely its origin are electrons deflected in the geomagnetic field and
emitting synchrotron radiation in the radio frequency range \cite{allanrev}.
Calculations show, that on ground level signals in the range of
$\sim\mu$V/(m\,MHz) are expected at frequencies of a few tens of MHz and
distances to the shower core smaller than 250~m \cite{huegefalcke}.
The LOPES experiment registers radio signals in the frequency range from 40 to
80~MHz \cite{lopesspie}. In this band are few strong man made radio
transmitters only, the emission from air showers is still strong (it decreases
with frequency), and background emission from the Galactic plane is still low.  
An active short dipole has been chosen as antenna. 
An inverted V-shaped dipole is positioned about 1/4 of the
shortest wavelength above an aluminum ground plate. In this way a broad
directional beam pattern is obtained.
Thirty antennas have been installed at the site of the KASCADE-Grande
experiment \cite{nehlspune}. LOPES is triggered on large air showers detected
with KASCADE-Grande.  All antennas, including the complete analog electronic
chain, have been individually calibrated with a reference radio source
\cite{nehlsarena}.

One of the most important results is the dependence of the radio signal pulse
height on the angle between the shower axis and the geomagnetic field
\cite{radionature,hornefferarena}. The measured radio pulse height normalized
to the number of muons in the respective shower is presented as function of the
cosine of the angle with respect to the geomagnetic field in \fref{lopes} ({\sl
left}). The signal yield clearly depends on the orientation with respect to the
magnetic field. This can be interpreted as confirmation for the proposed origin
of the radio frequency radiation, i.e.  synchrotron radiation in the
geomagnetic field.  It is planned to measure the polarization of the radio
signal as well, this will clarify the situation.  The measured radio pulse
height is shown as function of the registered number of muons in \fref{lopes}
({\sl right}).  The radio signal yield increases as function of muon number.
The latter is strongly correlated to the shower energy (nearly independent of
the mass of the primary), the range depicted corresponds roughly to about
$10^{17}$ to $6\cdot10^{17}$~eV.

In addition, alternative antenna designs are investigated \cite{gemmekearena}.
The experimental work is accompanied by efforts to include models for the radio
signal generation into the standard air shower simulation program CORSIKA
\cite{huegepune}.


\section*{References}

\end{document}